\documentclass{nature}

\usepackage[toc,page,title,titletoc,header]{appendix}
\usepackage{stmaryrd}
\SetSymbolFont{stmry}{bold}{U}{stmry}{m}{n}
\usepackage{amsmath}
\usepackage{amssymb}
\usepackage{amsthm}
\usepackage{tabularx}
\usepackage{subfigure}
\usepackage{epsfig}
\usepackage{indentfirst}
\usepackage{cases}
\usepackage{graphicx}
\usepackage{color}
\usepackage{float}
\usepackage{multirow}

\def\subFS{\scriptscriptstyle{FS}}
\def\subVS{\scriptscriptstyle{VS}}

\usepackage{graphicx}
\makeatletter
\let\saved@includegraphics\includegraphics
\AtBeginDocument{\let\includegraphics\saved@includegraphics}
\renewenvironment*{figure}{\@float{figure}}{\end@float}
\makeatother

\title{Power-law scaling for solid-state dewetting of thin films: an Onsager variational approach}

\author{Wei Jiang$^{1,2}$, Xianmin Xu$^{3,4}\footnote{email:xmxu@lsec.ac.cc.cn}$, Weizhu Bao$^5$ \& David J. Srolovitz$^{6}$\footnote{email:srol@cityu.edu.hk}}

\begin{document}

\maketitle

\begin{affiliations}
 \item School of Mathematics and Statistics, Wuhan University, Wuhan, 430072, China.
 \item Hubei Key Laboratory of Computational Science, Wuhan University, Wuhan, 430072, China.
 \item LSEC, ICMSEC, Academy of Mathematics and Systems Science, Chinese Academy of Sciences, Beijing 100190, China.
 \item School of Mathematical Sciences,University of Chinese Academy of Sciences, Beijing 100049, China.
 \item Department of Mathematics, National University of Singapore, Singapore, 119076.
 \item Departments of Materials Science and Engineering, City University of Hong Kong, Hong Kong SAR, China.
\end{affiliations}

\begin{abstract}
We examine the kinetics of surface diffusion-controlled, solid-state dewetting by consideration of the retraction of the contact in a semi-infinite solid thin film on a flat rigid substrate.
The analysis is performed within the framework of the Onsager variational principle applied to surface diffusion-controlled morphology evolution.
Based on this approach, we derive a simple, reduced-order model to quantitatively analyse the power-law scaling of the dewetting process.
Using asymptotic analysis and numerical simulations for the reduced-order model, we find that the  retraction distance grows as the $2/5$ power of time and the height of the ridge, adjacent to the contact, grows as the $1/5$ power of time for late time.
While the asymptotic analysis focuses on late time and a relatively simple geometric model, the Onsager approach is applicable to all times and descriptions of the morphology of arbitrary complexity.
\end{abstract}

\section{Introduction}

Solid-state dewetting of thin films on substrates has been observed in a wide range of systems by many research groups over many decades\cite{Thompson12,Jiran90,Pierre09b,Ye10b,Ye11b,Leroy16,Kovalenko17,
Naffouti17}.
While solid-state dewetting is deleterious, in the sense that it leads to the destabilization/agglomeration of continuous deposited films, it can also be exploited to produce a controlled distribution of particles on a substrate.
Recent examples include the formation of ordered arrays of nanoparticles and quantum dots, which have been exploited to produce sensors\cite{Armelao06, Mizsei93}, optical and magnetic devices\cite{Armelao06, Rath07} and for catalysts for the growth of carbon and semiconductor nanotubes and nanowires\cite{Randolph07, Schmidt09}.
Interest in such applications has driven research into the underlying mechanisms of solid-state dewetting\cite{Srolovitz86b,Wong00,Dornel06,Bussmann11,Dufay11,Jiang12,Rabkin14,Wang15,Zucker16b,Jiang16,Jiang18,Jiang19b}.

Dewetting of thin solid films is similar in many aspects to the wetting/dewetting of liquid thin films\cite{deGennes85}.
A major difference, however, is that in most applications, its mass transport is dominated by surface diffusion rather than fluid dynamics\cite{Jiang12,Wang15,Tripathi18}.
A typical feature of surface diffusion-controlled solid-state dewetting is the formation of a thickened ridge followed by a valley at the retracting film edge, the amplitudes of which increase with time/retraction distance\cite{Wong00,Dornel06,Ye10b,Ye11a,Kim13,Zucker13}.
Experimental observations show that the edge retraction distance scales as the $2/5$ power of time (at long time)\cite{Kim13,Zucker13}.
While this power-law has been widely observed in numerical simulations of solid thin film dewetting\cite{Wong00,Jiang12,Wang15}, rigorous theoretical analysis has remained elusive (despite reasonable approximate solutions\cite{Brandon66,Zucker16b}).

Brandon and Bradshaw (referred to as the BB model here) presented a simple geometric model for analysing this problem more than half a century ago\cite{Brandon66}.
Based on the observation that edge retraction and ridge growth are the main hallmarks of the experimental observation of surface diffusion-controlled solid state dewetting, BB described the cross-sectional profile of the discontinuous film as a semicircle that hits the substrate at the right contact angle (i.e., a contact angle of $90^\circ$) and connects with the film with a uniform thickness.
With this simple profile, they obtained two important scaling laws: the radius of a growing hole increases with the $2/5$ power of time, and the ridge height grows with the $1/5$ power of time.
The correspondence of these results with experimental observations demonstrate that this simple profile is sufficient to capture the essential features of surface diffusion-controlled solid-state dewetting.

Zucker {\it et al.} reexamined the BB model with the generalization that the cross-section of the profile needs not be semi-circular and the contact/Young angle $\theta_0$ needs not be $90^\circ$ (see Fig.~\ref{fig:closedSF}).
Their solution, for the mass-conserving (while the BB solution does not conserve the mass) surface diffusion-controlled dewetting, reproduces the $2/5$ power-law\cite{Zucker16b,Zucker15} of the retracting distance in the long-time limit.
However, their solution is both complicated and approximate.

In this paper, we present a new approach for analyzing the power-law scaling of
surface diffusion-controlled, solid-state dewetting that is both rigorous (conserving the total mass) and based upon the irreversible thermodynamics variational approach for surface diffusion-controlled morphology evolution problems\cite{Jiang19b}.
While this work builds on our earlier developments in the application of the Onsager variational principle to surface diffusion-controlled, capillarity-driven morphology evolution\cite{Jiang19b}, {\it{this paper specifically: (1) addresses the non-trivial power-law scaling of solid-state dewetting, and (2) provides an example of how to apply the Onsager approach to reduce the standard normal partial differential equations describing the morphology evolution to a set of ordinary differential equations that can be solved via asymptotic analysis and direct numerical solutions.}}

This paper is organized as follows. In the next section, we briefly review the application of the Onsager variational principle to construct a reduced-order model for an evolving, dissipative system. In Section 3, we apply this approach to the power-law retraction of the edge of a semi-infinite thin film on a substrate that results in an ordinary differential equation (ODE) for surface diffusion-controlled solid-state dewetting. In Section 4 and 5, we provide both asymptotic analysis and numerical simulations for the resulted ODE, and recover the $2/5$ experimentally and simulation-observed power-law. Finally, we draw some conclusions in Section 6.

\section{The Onsager variational principle}

The Onsager variational principle, formulated in 1931\cite{Onsager31a,Onsager31b}, is based on the reciprocal symmetry in linear irreversible thermodynamics.
This variational principle has found wide application in deriving evolution equations in fluid dynamics\cite{Qian06,Xu16,Man16,Qian17,Di18} and soft matter physics\cite{Doi11,Doi13book,Doi15}.
We apply this variational approach to surface-diffusion controlled, solid-state dewetting \cite{Jiang19b}.

Consider an isothermal system that may include interfaces  (e.g., the solid-vapor interface, solid-substrate interface, and vapor-substrate interface).
If the system deviates from its equilibrium state, then there will be spontaneous processes that tend to bring the system back to equilibrium.
In the linear response regime (i.e., not far from equilibrium), the time evolution of the system is governed by a variational principle.
Let $\alpha(t)=(\alpha_1(t),\alpha_2(t),\ldots,\alpha_n(t))$ be a set of state variables.
The time evolution of the system, may be described as the time derivatives of these state variables
$\dot{\alpha}(t)=(\dot{\alpha}_1(t),\dot{\alpha}_2(t),\ldots,\dot{\alpha}_n(t))$ (a raised dot ``$\cdot$'' denotes a time derivative);  it is determined by minimizing the ``Rayleighian'' $\mathcal{R}$ with respect to the rates $\{\dot{\alpha}_i\}$\cite{Doi15,Xu16,Suo97,Jiang19b}:
\begin{equation}
\mathcal{R}(\alpha,\dot{\alpha})=\dot{W}(\alpha, \dot{\alpha})+\Phi(\alpha,\dot{\alpha}).
\label{eqn:rayleigh}
\end{equation}
Here, $W(\alpha):= W(\alpha_1,\alpha_2,\ldots,\alpha_n)$ represents the total free energy of the system (a state function) and
$\dot{W}$ is the rate of change of $W$,
\begin{equation}
\dot{W}(\alpha, \dot{\alpha})=\sum_{i}\frac{\partial W}{\partial \alpha_i}\dot{\alpha_i}.
\end{equation}

$\Phi(\alpha,\dot{\alpha})$, in Eq.~\eqref{eqn:rayleigh}, is the free energy dissipation function; it is half the free energy dissipation rate.
In the linear response regime, the dissipation function can be written as a quadratic function of the rates $\{\dot{\alpha}_i\}$; i.e.,
\begin{equation}
\Phi(\alpha,\dot{\alpha})=\frac{1}{2}\sum_{i,j}\zeta_{ij}(\alpha)\dot{\alpha}_i\dot{\alpha}_j,
\end{equation}
where the damping/friction coefficients $\zeta_{ij}$ form a symmetric,  positive definite matrix.
Minimizing the Rayleighian with respect to the rates $\{\dot{\alpha}_i\}$ yields a set of kinetic equations
\begin{equation}
\sum_{j}\zeta_{ij}\dot{\alpha}_j=-\frac{\partial W}{\partial \alpha_i},\qquad i=1,2,\ldots,n.
\end{equation}
This describes the force balance between the potential force $-\frac{\partial W}{\partial \alpha_i}$ and the dissipative force $\frac{\partial \Phi}{\partial \dot{\alpha}_i}$ (which is linear in the rates $\{\dot{\alpha}_i\}$).
A simple calculation shows that the variational principle leads to $\dot{W}=-2\Phi$, which means $\Phi$ is half the rate of free energy dissipation, as asserted above.
This variational principle for {\it isothermal} systems, outlined above, can be generalized to {\it non-isothermal} systems via  the maximization of the Onsager-Machlup action\cite{Onsager31a,Onsager31b}.

The evolution of a dissipative system described by a set of field variables can be approximated by a {\it finite} set of state variables.
The total free energy $W$ and dissipation function $\Phi$ can be obtained as functions of these state variables and their time derivatives.
Application of the Onsager variational principle then gives a system of {\it ordinary differential equations} (ODEs) that describes the time evolution of the state variables, i.e., the time evolution of the system~\cite{Xu16,Man16,Doi15}.
In our Onsager variational approach to the surface diffusion-controlled, solid-state dewetting problem, we can consider the field variables as a finite set of variables that represent a reduced-order description of the film profile. The purpose of the present work is to apply the Onsager's variational principle to the dynamics of solid-state dewetting.
While it is possible to use this approach to accurately describe the evolution of the entire film profile, we focus here on deriving the power law of the retraction of the thin film dewetting front for which a relative simple description of the film profile suffices.
We derive a reduced model for the dynamics of the film by using the Onsager variational principle, then perform asymptotic analysis and numerical simulations to derive the power law scaling of the retraction front in the long-time limit.

\section{A reduced-order variational model}

\begin{figure}
\centering
\includegraphics[width=11cm,angle=0]{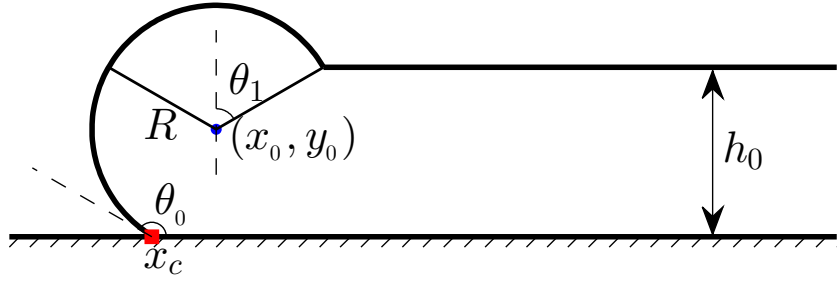}
\caption{A schematic illustration of the model of the retracting contact and the correspond ridge.
We assume that the profile of the film-vapor interface consists of a circular arc of  radius  $R:=R(t)$ (center  located
at  $(x_0(t), y_0(t)$)), and a straight line representing the semi-infinite film of thickness $h_0$, measured from the substrate (i.e., the $x$-axis).  The circular arc meets the substrate at the isotropic Young's angle $\theta_0$.  For convenience, all lengths are scaled by the film thickness.}
\label{fig:closedSF}
\end{figure}
We assume the film geometry with a retracting contact and ridge  proposed by Zucker {\it et al.}\cite{Zucker16b}, as shown in Fig.~\ref{fig:closedSF}.
We assume that the contact satisfies the  isotropic Young's angle $\theta_0$, i.e., $\cos\theta_0 = (\gamma_{\subVS}-\gamma_{\subFS})/\gamma_0$, where $\gamma_0$, $\gamma_{\subVS}$, $\gamma_{\subFS}$ represent the film-vapor, vapor-substrate and film-substrate interface energy per unit length.
For simplicity in notations, we also introduce an angle $\theta_1(t)$ to characterize the position of the other end of the curve.
We further assume that no free energy is dissipated by the motion of the contact point, i.e., no contact drag (this is consistent with the assumption that the contact angle always satisfies the equilibrium Young's angle condition).

In our application of the Onsager variational principle, we focus on the reduced order model where the film profile is characterized by
the parameters $R(t)$, $x_0(t)$, $y_0(t)$ and $\theta_1(t)$ that evolve during the dewetting process.
These parameters are not independent; they satisfy several geometric constraints such that there is only one independent variable.
Examination of the geometry of Fig.~\ref{fig:closedSF} shows that
\begin{equation*}
\left\{
\begin{array}{lll}
R\cos(\pi-\theta_0)&=&y_0, \\[0.4em]
y_0+R\cos\theta_1&=&h_0.
\end{array}
\right.
\end{equation*}
This implies that both $y_0$ and $\theta_1$ can be written as functions of $R$.
These equations imply
\begin{equation}\label{eqn:theta1}
R(\cos\theta_1-\cos\theta_0)=h_0.
\end{equation}
Taking the time derivative of both sides of this equation, yields
\begin{equation}
(\cos\theta_1-\cos\theta_0)\,\dot{R}=(R\sin\theta_1)\,\dot{\theta}_1,
\label{eqn:IM1}
\end{equation}
which implies
\begin{equation}
\dot{\theta}_1=g_1(R,\theta_0,\theta_1)\,\dot{R}, \quad \text{where}\quad g_1(R,\theta_0,\theta_1)=\frac{\cos\theta_1-\cos\theta_0}{R\sin\theta_1}.
\label{eqn:IMEX1}
\end{equation}

Next, we consider the conservation of the film mass.
The area $A:=A(t)$ of the thin film enclosed by the circular arc and the rigid substrate is
\begin{equation}
A=\frac{1}{2}R^2(\theta_0+\theta_1)-\frac{1}{2}R^2\sin\theta_0\cos\theta_0+
\frac{1}{2}R^2\sin\theta_1\cos\theta_1-R^2\cos\theta_0\sin\theta_1.
\label{eqn:formA}
\end{equation}
The conservation of thin film mass implies
\begin{equation}
\frac{{\rm d} A}{{\rm d} t}=\frac{\rm d}{{\rm d} t}\bigl[h_0(x_0+R\sin\theta_1)\bigr].
\label{eqn:masscon}
\end{equation}
Inserting Eq.~\eqref{eqn:formA} into the above equation, we obtain a second important geometric relation
\begin{eqnarray}
h_0 \dot{x}_0&=&\Bigl[R\bigl(\theta_0+\theta_1-\frac{1}{2}\sin 2\theta_0+\frac{1}{2}\sin 2\theta_1-2\cos\theta_0\sin\theta_1\bigr)-h_0\sin\theta_1\Bigr]\dot{R}  \nonumber \\[0.5em]
&&+\Bigl[\frac{1}{2}R^2(1-2\cos\theta_0\cos\theta_1+\cos 2\theta_1)-h_0 R\cos\theta_1 \Bigr] \dot{\theta}_1.
\label{eqn:IM2}
\end{eqnarray}
From the above relation, and making use of Eq.~\eqref{eqn:IMEX1}, we obtain
\begin{equation}
\dot{x}_0=g_2\,\dot{R},
\label{eqn:IMEX2}
\end{equation}
where $g_2$ is given as
\begin{eqnarray*}
g_2&=&\Bigl[\frac{R}{h_0}\bigl(\theta_0+\theta_1-\frac{1}{2}\sin 2\theta_0+\frac{1}{2}\sin 2\theta_1-2\cos\theta_0\sin\theta_1\bigr)-\sin\theta_1\Bigr] \\[0.5em]
&&+\Bigl[\frac{R}{2h_0}(1-2\cos\theta_0\cos\theta_1+\cos 2\theta_1)-\cos\theta_1 \Bigr]\,\frac{\cos\theta_1-\cos\theta_0}{\sin\theta_1}.
\end{eqnarray*}
Substituting Eq.~\eqref{eqn:theta1} into this expression for  $g_2$, we obtain  $g_2$ as a function of  $\theta_0$ and $\theta_1$,
\begin{equation}
g_2:=g_2(\theta_0,\theta_1)=\frac{1}{\cos\theta_1-\cos\theta_0}(\theta_0+\theta_1-\sin\theta_0\cos\theta_0-\sin\theta_1\cos\theta_0).\label{eqn:g2}
\end{equation}

With the above geometrical relations in hand, we derive a reduced model for describing the dynamics of the dewetting process by application of the Onsager variational principle.
We choose $R(t)$ as the only free variable.
The total interfacial free energy of this system $W:=W(t)$ can be written as
\begin{equation}
W=\gamma_0\bigl[R(\theta_0+\theta_1)-(x_0+R\sin\theta_1)\bigr]+\gamma_0\cos\theta_0(x_0-R\sin\theta_0).
\label{eqn:totalenergy}
\end{equation}
Taking its time derivative and making use of Eqs.~\eqref{eqn:IMEX1} and
\eqref{eqn:IMEX2}, we have
\begin{equation}
\dot{W}=\frac{\partial W}{\partial R}\dot{R},
\end{equation}
where
\begin{equation}
\frac{\partial W}{\partial R}=\gamma_0\Bigl[(\theta_0+\theta_1-\sin\theta_1-\cos\theta_0\sin\theta_0)+R(1-\cos\theta_1)g_1-(1-\cos\theta_0)g_2\Bigr].
\label{eqn:DEng}
\end{equation}

Next, we compute the energy dissipation function for the evolving profile.
For this, we parameterize the circular section of the  film/vapor profile as
\begin{equation}
\left\{
\begin{array}{l}
x(\theta,t)=x_0(t)+R(t)\sin\theta,   \\ [0.6em]
y(\theta,t)=R(t)(\cos\theta-\cos\theta_0),
\end{array}
\right.
\end{equation}
where $\theta \in [-\theta_0, \theta_1]$.
The procedure for obtaining the dissipation function $\Phi$ is  similar to that presented previously~\cite{Jiang19b}.
We write the normal velocity of the interface curve $v_n(\theta)$ as
\begin{equation}
v_n(\theta)=\dot{x}_0\sin\theta+\dot{R} (1-\cos\theta_0\cos\theta),
\quad  \theta \in [-\theta_0, \theta_1].
\end{equation}
The corresponding (mass) flux $j:=j(\theta)$, $\theta \in [-\theta_0, \theta_1]$ along the circular arc is
\begin{eqnarray}
j(\theta)&=&\int_{-\theta_0}^{\theta}v_n(\theta)R{\rm d}\theta \nonumber\\[0.6em]
&=&-R(\cos\theta-\cos\theta_0)\dot{x}_0+R(\theta+\theta_0-\sin\theta\cos\theta_0-\sin\theta_0
\cos\theta_0)\dot{R}\nonumber\\
&=&g_3(\theta,\theta_0,\theta_1)R\dot{R},
\label{eqn:flux}
\end{eqnarray}
where 
\begin{eqnarray}
g_3(\theta,\theta_0,\theta_1)&=&\Bigl[-(\cos\theta-\cos\theta_0)\Bigl(\theta_0+\theta_1-(\sin\theta_0+\sin\theta_1)
\cos\theta_0\Bigr)\nonumber\\
&&+(\cos\theta_1-\cos\theta_0)\Bigl(\theta+\theta_0-(\sin\theta+\sin\theta_0)
\cos\theta_0\Bigr) \Bigr]\frac{1}{\cos\theta_1-\cos\theta_0}\label{eqn:g3}.
\end{eqnarray}
Here, we employed Eqs.~\eqref{eqn:IMEX2} and \eqref{eqn:g2} and imposed the zero-mass flux boundary condition at the contact point
$j(-\theta_0)=0$ (this implies that the total area/mass is conserved during the evolution)\cite{Jiang19b,Wang15}.

The dissipation function $\Phi$ can  be written as\cite{Jiang19b}
\begin{equation}
\Phi=\frac{1}{2}\frac{k_BT}{D_s\nu\Omega_0^2}\int_{-\theta_0}^{\theta_1}j^2(\theta)R\,{\rm d}\theta,
\end{equation}
where $D_s$ is the surface diffusivity, $\nu$ is the number of diffusing atoms per unit area, $\Omega_0$ is the atomic volume, and $k_BT$ is the thermal energy.
By inserting~\eqref{eqn:flux} into this expression and making use of~\eqref{eqn:IMEX2},
it can be recast into the following quadratic form with respect to the rate function $\dot{R}$
\begin{equation}
\Phi:=\Phi(R,\dot{R})=\frac{1}{2}\zeta(R,\theta_0,\theta_1)\dot{R}^2,
\end{equation}
where the friction coefficient $\zeta:=\zeta(R,\theta_0,\theta_1)$ is
\begin{equation}
\zeta(R,\theta_0,\theta_1)=\frac{k_BT}{D_s\nu\Omega^2}R^3\int_{-\theta_0}^{\theta_1}g_3(\theta,\theta_0,\theta_1)^2\,{\rm d}\theta.
\label{eqn:zeta}
\end{equation}

Applying the Onsager variational principle\cite{Doi13book,Jiang19b}, we write the Rayleighian of our system in terms of the free energy $W:=W(R)$ and dissipation function $\Phi:=\Phi(R,\dot{R})$:
\begin{equation}
\mathcal{R}(R,\dot{R})=\frac{\partial W}{\partial R}\,\dot{R}+\Phi(R,\dot{R}).
\end{equation}
Minimization of the Rayleighian $\mathcal{R}$ with respect to the rate variable $\dot{R}$ yields the following
evolution equation for the radius function $R:=R(t)$,
\begin{equation}
\zeta(R,\theta_0,\theta_1)\dot{R}=-\frac{\partial W}{\partial R},
\label{eqn:ODE}
\end{equation}
where the function $\theta_1:=\theta_1(t)$ is updated according to Eq.~\eqref{eqn:theta1}.
This ODE~\eqref{eqn:ODE} governs the interface evolution of a retracting semi-infinite thin film depicted by Fig.~\ref{fig:closedSF}; this is a reduced-order variational model for the dewetting  of a solid film on a substrate  via surface diffusion.
An alternative approach is to solve the coupled ODEs~\eqref{eqn:ODE} and \eqref{eqn:IMEX1} with respect to $R$ and $\theta_1$ to obtain the interface evolution.

\section{Asymptotic analysis}

We first perform an asymptotic analysis of the ODE (i.e., Eq.~\eqref{eqn:ODE}) to obtain a simple, power-law description of dewetting.
For simplicity of presentation, we focus on the special case of  $\theta_0={\pi}/{2}$. 
Numerical results are presented for other Young's angles below.
For $\theta_0={\pi}/{2}$, the expressions for  ${\partial W}/{\partial R}$ and  $\zeta(R,\theta_0,\theta_1)$ in Eq.~\eqref{eqn:ODE} can be simplified.
Inserting Eqs.~\eqref{eqn:IMEX1} and \eqref{eqn:g2} into Eq.~\eqref{eqn:DEng} gives
\begin{equation}\label{eqn:tempW}
-\frac{\partial W}{\partial R}=-\gamma_0\Bigl\{\frac{1}{\cos\theta_1}\Bigl[(\frac{\pi}{2}+\theta_1)(\cos\theta_1-1)-
\sin\theta_1\cos\theta_1\Bigr]+\frac{1}{\sin\theta_1}(1-\cos\theta_1)\cos\theta_1\Bigr\}.
\end{equation}
The expression $\zeta(R,\theta_0,\theta_1)$ in Eq.~\eqref{eqn:zeta} simplifies by rewriting
$g_3= -({\pi}/{2}+\theta_1)\cos\theta/\cos\theta_1+\theta+{\pi}/{2}$
and
\begin{eqnarray}
\int_{-\theta_0}^{\theta_1}g_3(\theta,\frac{\pi}{2},\theta_1)^2\,{\rm d}\theta
&=&\frac{1}{2}\Bigl(\frac{\theta_1 +{\pi}/{2}}{\cos\theta_1}\Bigr)^2
\Bigr(\frac{\pi}{2}+\theta_1+\sin \theta_1 \cos \theta_1\Bigl)+\frac{1}{3}\Bigr(\theta_1+\frac{\pi}{2}\Bigl)^3 \nonumber \\
&&-2\Bigl(\frac{\theta_1 +{\pi}/{2}}{\cos\theta_1}\Bigr)\Bigr(\frac{\pi}{2}\sin\theta_1+\theta_1\sin\theta_1+\cos\theta_1\Bigl).
\label{eqn:tempGamma}
\end{eqnarray}

In the long-time limit, we can assume $R\gg h_0$ such that $\cos \theta_1 = {h_0}/{R}\ll 1$.
This implies that $\theta_1\approx {\pi}/{2}+\mathcal{O}({h_0}/{R})$
and $\sin\theta_1\approx 1+\mathcal{O}(({h_0}/{R})^2)$.
With these approximations, Eq.~\eqref{eqn:tempW} reduces to
\begin{eqnarray}
-\frac{\partial W}{\partial R}
&=&-\gamma_0\Bigl[ \Bigl(\frac{R}{h_0}\Bigr)     \Bigl( \Bigl[\pi + \mathcal{O}\Bigl(\frac{h_0}{R}\Bigr)\Bigr]
\Bigl(\frac{h_0}{R}-1\Bigr) -
\Bigl[1+\mathcal{O}\Bigl(\frac{h_0}{R}\Bigr)^2\Bigr]\frac{h_0}{R}\Bigr)+
\frac{1-{h_0}/{R}}{1+\mathcal{O}(({h_0}/{R})^2)}\Bigl(\frac{h_0}{R}\Bigr)\Bigr]\nonumber\\
&\approx& \pi\gamma_0\Bigl(\frac{R}{h_0}\Bigr),
\label{eqn:rhs}
\end{eqnarray}
where in the last line we keep only the leading order term.
In the same limit, Eq.~\eqref{eqn:tempGamma}  reduces to
\begin{eqnarray*}
\int_{-\theta_0}^{\theta_1}g_3(\theta,\theta_0,\theta_1)^2\,{\rm d}\theta
&=&\frac{1}{2}\Bigl[{\pi+\mathcal{O}\Bigl(\frac{h_0}{R}\Bigr)}^2\Bigr]
\Bigl({\frac{R}{h_0}}\Bigr)^{2}
\Bigl(\pi+\mathcal{O}\Bigl(\frac{h_0}{R}\Bigr)
+\Bigl[1+\mathcal{O}\Bigr(\Bigr(\frac{h_0}{R}\Bigl)^2\Bigl)\Bigl]\frac{h_0}{R}\Bigr)\nonumber\\
&&-2\Bigl[{\pi+\mathcal{O}\Bigl(\frac{h_0}{R}\Bigr)}\Bigr] \Bigl({\frac{R}{h_0}}\Bigr)
\Biggl(\frac{\pi}{2}\Bigl[1+\mathcal{O}\Bigr(\Bigr(\frac{h_0}{R}\Bigl)^2\Bigl)\Bigl]\nonumber\\
&&+\Bigl[\frac{\pi}{2}+\mathcal{O}\Bigl(\frac{h_0}{R}\Bigr)\Bigr]\Bigl(1+\mathcal{O}\Bigr(\Bigr(\frac{h_0}{R}\Bigl)^2\Bigr)+\frac{h_0}{R}\Biggr)
+\frac{1}{3}\Bigl(\pi+ \mathcal{O}(\frac{h_0}{R})\Bigr)^3\nonumber\\
&\approx& \frac{\pi^3 }{2}\Bigl(\frac{R}{h_0}\Bigr)^2.
\label{eqn:g3int}
\end{eqnarray*}
Inserting this expression into Eq.~\eqref{eqn:zeta}, we  obtain
\begin{equation*}
\zeta(R,\frac{\pi}{2},\theta_1)\approx \frac{\pi^3 k_BT}{2D_s\nu\Omega_0^2h_0^2}R^5.
\end{equation*}
Finally, by inserting this expression and Eq.~\eqref{eqn:rhs} into the ODE describing dewetting Eq.~\eqref{eqn:ODE}, we obtain a simplified (leading-order) ODE for dewetting:
\begin{equation}
R^4 \dot{R}=a,
\label{eqn:simpleODE}
\end{equation}
where $a={2B\gamma_0h_0}/{\pi^2}$ and $B={D_s\nu\Omega_0^2}/{k_BT}$ is a material constant.

Eq.~\eqref{eqn:simpleODE} demonstrates that
\begin{equation*}
R(t) = (5at+C_1)^{1/5}\propto (5a)^{1/5}\, t^{{1}/{5}},
\end{equation*}
where $C_1$ is a constant that depends on the initial value of $R$.
This implies that the ridge adjacent to the moving contact grows with time in a power law fashion as $t^{1/5}$.
This power-law exponent is consistent with  previous analysis\cite{Brandon66,Zucker16b} and experiment.

Next, we examine how the contact point $x_c:=x_c(t)$ moves at long time.
The contact point evolution is related to $R(t)$ by
\begin{equation}
x_c(t)=x_0(t)-R(t)\sin\theta_0=x_0(t)-R(t),
\end{equation}
then by taking the time derivative and using Eq.~\eqref{eqn:IMEX2}, we obtain
\begin{equation}
\dot{x_c}=(g_2-1)\dot{R}.
\label{eqn:CTP}
\end{equation}
Eq.~\eqref{eqn:g2} implies
\begin{equation}
g_2-1=\frac{1}{\cos\theta_1}\Bigl(\frac{\pi}{2}+\theta_1\Bigr)-1= \Bigl(\frac{h_0}{R}\Bigr)^{-1}\Bigl(\pi+\mathcal{O}\bigl(\frac{h_0}{R}\bigr) \Bigr)-1 \approx \pi\Bigl(\frac{h_0}{R}\Bigr)^{-1}= \frac{\pi R}{h_0},
\end{equation}
such that, in the long-time limit (i.e., to  leading-order),
\begin{equation}
\dot{x_c} = \frac{\pi R}{h_0}\dot{R}.
\label{e:CTP}
\end{equation}
Integration of this expression leads to
\begin{equation}
x_c(t)= \frac{\pi R^2(t)}{2h_0}+C_2\propto b\, t^{{2}/{5}},
\end{equation}
where $C_2$ is also a constant determined by the initial location of the contact point and
 $b^5={(5B\gamma_0)^2\pi}/{(2h_0)^{3}}$.
This is consistent with earlier results for the power-law dependence of the retraction distance with time in surface diffusion-controlled, solid-state dewetting\cite{Wong00,Jiang12,Kim13,Wang15,Zucker16b}.


\section{Numerical results}

The coupled ODEs, Eqs.~\eqref{eqn:ODE}-\eqref{eqn:IMEX1}, can be solved numerically for all times. We employ the classical fourth-order Runge-Kutta method for any initial conditions and Young's angle, $\theta_0$.
Here, because our main focus is on the power-law scaling of surface diffusion-controlled, solid state dewetting kinetics, for simplicity, we examine the initial values (e.g., $R(0)=2h_0$ and $\theta_1(0)=\pi/16$) under different isotropic Young's angles.
Note that although the initial values of $R$ and $\theta_1$ may not be consistent with Eq.~\eqref{eqn:theta1}, they will quickly become consistent after a short time of evolution.

\begin{figure}
\centering
\includegraphics[width=10cm,angle=0]{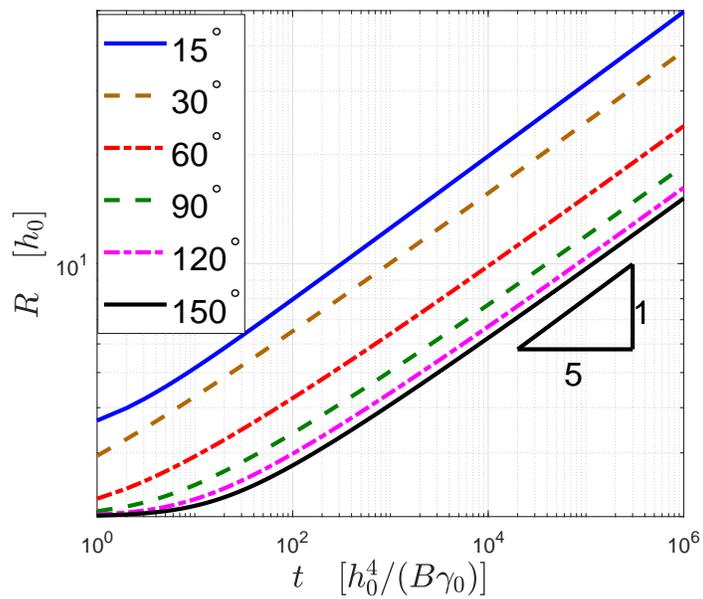}
\caption{The ridge radius $R$ (i.e., proportional to ridge height) versus time $t$ for six different Young's angles $\theta_0=15^\circ$, $30^\circ$, $60^\circ$, $90^\circ$, $120^\circ$, and $150^\circ$. The late time data is consistent with a power-law of the form $R\propto t^{1/5}$ for all Young's angles.}
\label{fig:radius}
\end{figure}

\begin{figure}
\centering
\includegraphics[width=10cm,angle=0]{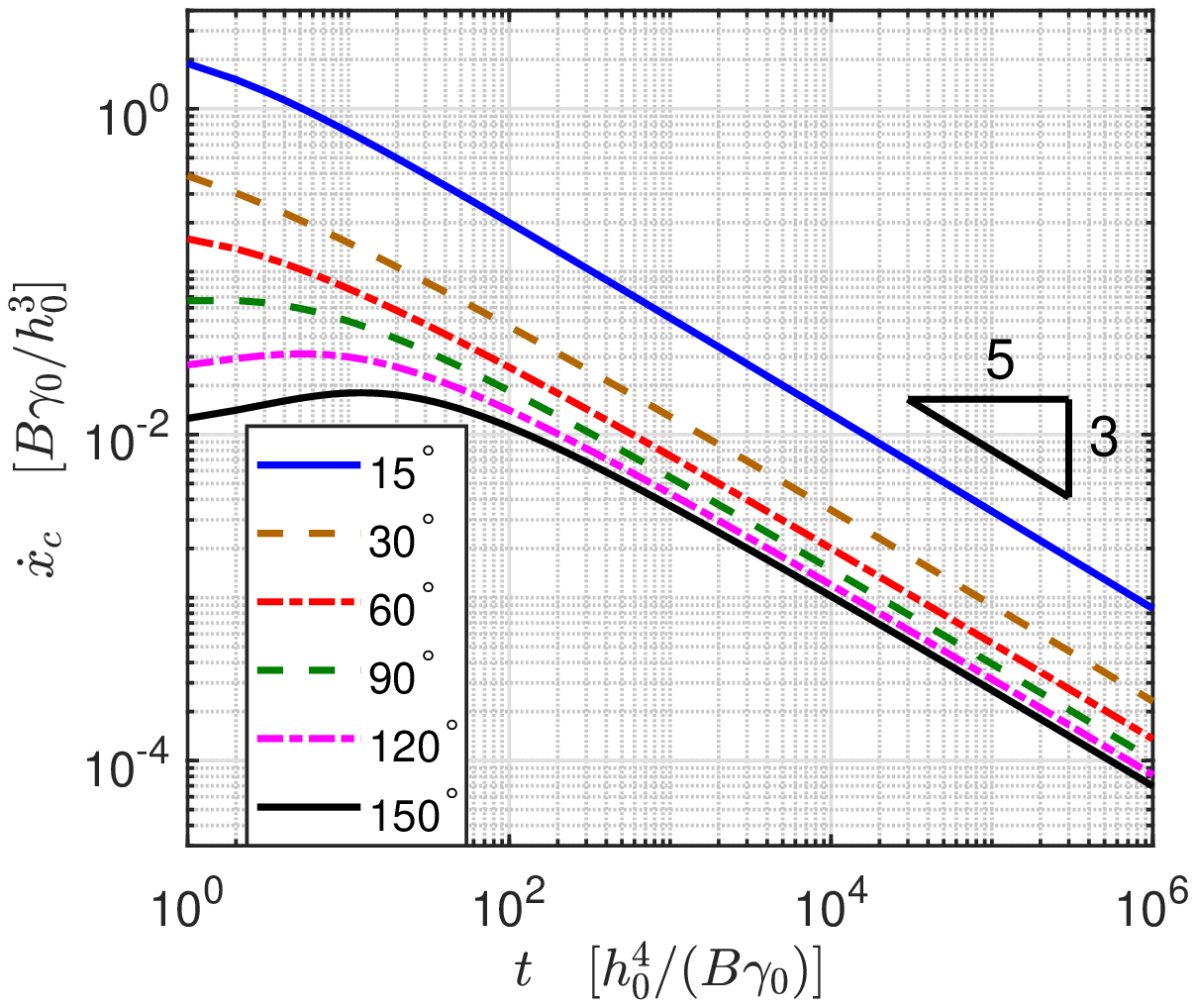}
\caption{The retraction velocity of the contact point $\dot{x}_c$ {\it versus} time $t$ for the same six Young's angles as in Fig.~\ref{fig:radius}, where a power law that $\dot{x}_c\propto t^{-3/5}$ is clearly shown. The late time data is consistent with a power-law of the form $x_c\propto t^{2/5}$ for all Young's angles.}
\label{fig:num}
\end{figure}

Figures~\ref{fig:radius} and \ref{fig:num} show numerical results for the evolution of the ridge radius and contact retraction velocity for six different isotropic Young's angles $\theta_0=15^\circ$, $30^\circ$, $60^\circ$, $90^\circ$, $120^\circ$ and $150^\circ$.
As shown in Fig.~\ref{fig:radius}, a log-log plot of the ridge radius $R$ {\it versus} time clearly exhibits a $1/5$ power-law at late times for all six different Young's angles. The ridge height (i.e., $R+y_0=R(1-\cos\theta_0)$) also follows the same power-law during  dewetting.
Figure~\ref{fig:num} shows that the contact retraction velocity $\dot{x}_c$ is consistent with the predicted power-law $\dot{x}_c\propto t^{-3/5}$ at late times, which indicates that the retraction distance $x_c(t)$ satisfies a $2/5$ power-law with time.

\section{Conclusions}

We examine the power-law scaling of the surface diffusion-controlled, solid-state dewetting of a semi-infinite thin film on a flat substrate.
Our approach is based upon the Onsager variational principle and motivated by earlier, simple geometric models that are consistent with the power-law retraction behavior observed in more complete numerical studies of the evolving film profile during solid-state dewetting.
The simplified nature of the film geometry, allows us to derive a reduced-order variational model, the evolution of which is governed by an ODE.
Asymptotic analysis and numerical simulations of the ODE reproduced the $2/5$ power-law of the retraction distance with time, and the $1/5$ time exponent for the height of the ridge adjacent to the moving contact line.
Although these power-laws have been predicted previously based on analysis and numerical simulation, the present results demonstrate the simplicity and applicability of the Onsager variational principle to describe surface diffusion-controlled morphology evolution problems in materials science. While the geometric model employed here is very simple, the Onsager variational principle approach is applicable to much more complex representations of the geometry; such generalizations yield a more complicated system of ODEs.

\begin{addendum}
 \item[Data Availability] The data supporting the findings of this study are available from the authors upon reasonable request.
 \item[Code Availability] The custom Matlab code used in numerical simulations is available from the authors upon reasonable request.
\end{addendum}

\section*{References}
\bibliography{thebib}


\begin{addendum}
 \item This work was partially supported by the National Key R\&D Program of China under Grant 2018YFB0704304 and Grant 2018YFB0704300(X.X.),the National Natural Science Foundation of China No.~11871384 (W.J.) and No.~11971469(X.X.), Natural Science Foundation of Hubei Province No.~2019CFA007 (W.J.) and the Ministry of Education of Singapore grant R-146-000-296-112 (W.B.). Part of the work was done when the authors were
     visiting the Institute for Mathematical Sciences
     at the National University of Singapore in 2020.
 \item[Author Contribution] W.J., W.B and D.J.S conceived the initial project, and W.J. and X.X. conducted the research. The manuscript was written through contributions of all authors. All authors have given approval to the final version of the manuscript.
 \item[Competing Interests] The authors declare that they have no competing interests.
 \item[Correspondence] Correspondence and requests for materials should be addressed to X.X. or D.J.S..
\end{addendum}


\end{document}